\documentclass[conference]{IEEEtran}
\IEEEoverridecommandlockouts
\usepackage{cite}
\usepackage{amsmath,amssymb,amsfonts}
\usepackage[ruled,longend,boxruled]{algorithm2e}
\usepackage{graphicx}
\usepackage{textcomp}
\usepackage{xcolor}
\def\BibTeX{{\rm B\kern-.05em{\sc i\kern-.025em b}\kern-.08em
    T\kern-.1667em\lower.7ex\hbox{E}\kern-.125emX}}
\begin{document}

\title{Concurrent Deterministic Skiplist and Other DataStructures\\
}

\author{\IEEEauthorblockN{Aparna Sasidharan}
\IEEEauthorblockA{\textit{Illinois Institute of Technology, RCCS Riken} \\
Chicago,IL,USA \\
Kobe, Japan\\
aparnasasidharan2017@gmail.com}
}


\maketitle

\begin{abstract}
	Skiplists are used in a variety of applications for storing data subject to order criteria. In this article we discuss the design, analysis and performance of a concurrent deterministic skiplist on many-core NUMA nodes~\cite{numa}. We also evaluate the performance of a concurrent lock-free unbounded queue implementation and two concurrent multi-reader, multi-writer~(MWMR) hash table implementations and compare them with those from Intel's Thread Building Blocks~(TBB) library~\cite{TBB}. We introduce strategies for memory management that reduce page faults and cache misses for the memory accesses in these data structures. This paper proposes hierarchical design and usage of concurrent data structures in programs to improve memory latencies by reducing memory accesses to remote NUMA nodes. 

\end{abstract}

\section{Motivation\label{motiv}}

Most modern high performance machines have many-core CPUs with high core counts~\cite{top500}. The CPUs may be used alongside accelerators for compute intensive applications. Many computational science applications have shown scalability~\cite{astro} on many-core and heterogeneous nodes due to their regular memory access patterns which have natural spatial and temporal localities, lower page faults and cache misses. Distributed data-intensive workloads~\cite{ycsb} which perform point location and range searches have less spatial and temporal localities compared to applications from computational science. These workloads do not scale as well as their counterparts in computational science on high performance machines. In this paper, we discuss the performance of three fundamental data structures that are used for storing and querying data in data-intensive applications. Performance of such applications on many-core NUMA nodes~\cite{numa} can be improved by using scalable concurrent data structures. For example, using suitable functions that map queries to keys, point location and range searches can be made parallel using concurrent hash tables and skiplists.   

This article focuses on a concurrent deterministic skiplist. Skiplists can function as ordered sets and are building blocks for ordered multi-sets and heaps. Concurrent randomized skiplist implementations contain nodes with random number of links~(heights)~\cite{herlihy}. The costs of insertion, search and deletion are $O(logn)$ with high probability. The $O(logn)$ computational complexity depends on the random number generator used by the implementation for generating node heights. The random number generator should generate nodes of height $j+1$ with probability $\frac{1}{t}^{j}$ where $t$ is the number of nodes to skip at level $1$ and $j+1$ is the level, if levels are numbered from $1$ to $L$. If probabilities follow this distribution, there will be $\frac{1}{t}^{j}*N$ links per level $j+1$ and all randomized skiplist operations will be $O(logn)$ with high probability~\cite{sedgewickbook}. 
A deterministic skiplist is balanced and its costs for insertion, find and deletion are $O(logn)$. Our implementation is a concurrent version of 1-2-3-4 trees~(deterministic skiplists)~\cite{sedgewick}. The skiplist has $log(N)$ levels and the number of links at each level is at least $\frac{1}{4}$ the links at the previous lower level. Levels are numbered from bottom~($0$) to root~($logn$). Leaf nodes are the links at level $1$. Leaf nodes contain pointers to the nodes of a linked-list which stores keys and data~(terminal nodes). The algorithms are discussed in detail in the remaining sections, along with performance.

Dictionaries or Hash Tables are useful data structures for organizing data based on keys, where the key universe is much larger than the number of keys used in a program~\cite{Knuth1973}. Operations on hash tables such as insert, delete and find can be performed in constant time per key. But the performance of concurrent hash tables is affected by random memory accesses which cause cache misses and page faults in programs which use large workloads. This paper discusses the performance of three implementations of concurrent multi-writer multi-reader~(MWMR) hash tables.

Queues are widely used for load balancing workloads, storing data and for communication. A poor queue implementation can affect the scalability of a program. This article discusses controlled memory allocation and recycling to reduce page faults in data structures using concurrent lock-free queues as a motivating example.

All experiments in this paper were performed on the Delta Supercomputer housed at the National Center for Supercomputing Applications~(NCSA)~\cite{delta} which has AMD Milan many-core NUMA nodes~\cite{AMD} and the Fugaku supercomputer at Riken~\cite{fugaku} with Fujitsu A64FX~\cite{fujitsu} nodes. The paper is organized as follows. Section~\ref{skiplist} discusses the design and analysis of the concurrent deterministic skiplist. Sections~\ref{queue}, \ref{queueperf} and \ref{mem} discuss concurrent queue implementations, their performance alongwith memory management in many-core processors. Scalability of concurrent skiplists is discussed in section~\ref{skiplistperf}. Concurrent hash tables and their performance are discussed in sections~\ref{ht} and \ref{htperf}. Sections~\ref{rel} and \ref{conc} have the related work, conclusions and future work.

\section{Skiplist\label{skiplist}}

\begin{figure}[!htbp]
\begin{center}
\includegraphics[width=3.4in,height=2.8in]{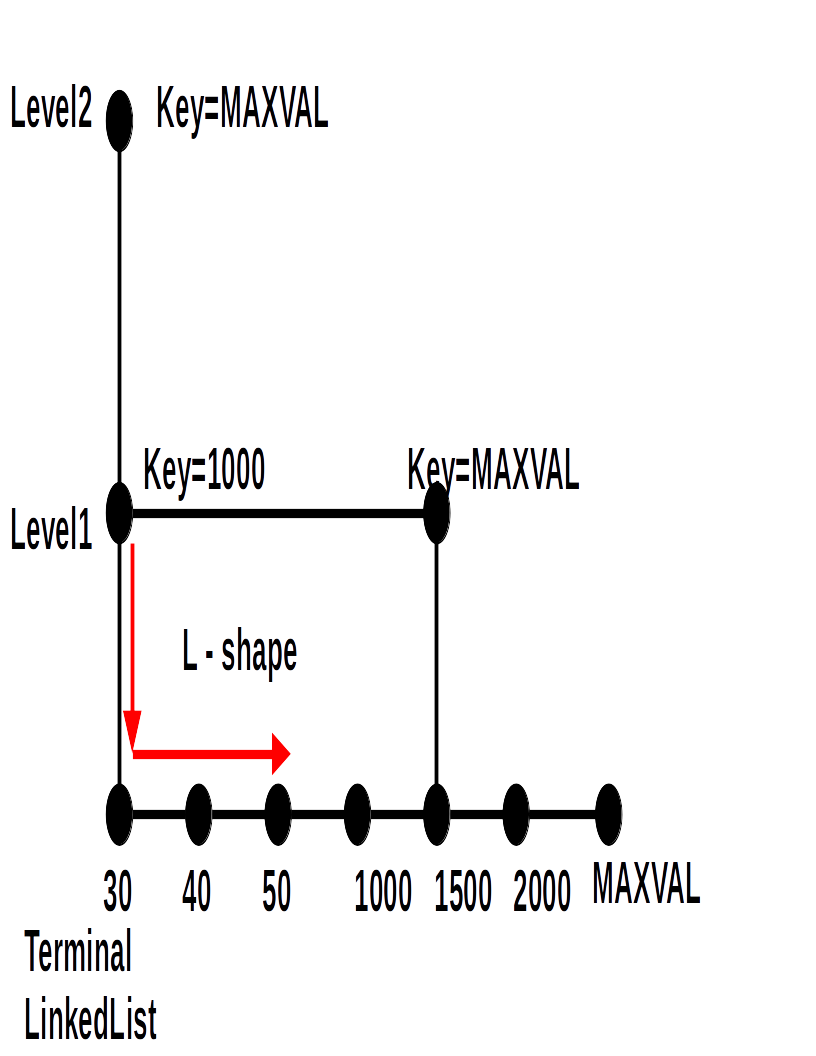}
\caption{\label{fig:tree}Deterministic Skiplist with 3 levels}
\end{center}
\end{figure}

In this section we discuss a concurrent 1-2-3-4 tree/skiplist based on the sequential version from Munro and Sedgewick~\cite{sedgewick}. A skiplist is an ordered linked-list with additional links. The extra links reduce search costs by providing shortcuts to different intervals in the linked-list. Without these links, search costs can be as high as the length of the list $O(n)$. In a random skiplist~\cite{randomskiplist}, lookup nodes divide the linked-list into non-overlapping intervals of random lengths. Although interval lengths are random, the total number of links traversed to find a node in a list with $n$ nodes is $O(logn)$ with high probability. A deterministic 1-2-3-4 skiplist divides the linked-list into intervals of at least $2$ nodes~(1 interval) and at most $5$ nodes~(4 contiguous intervals), with a hierarchy of links to these intervals. The data structure is described in figure~\ref{fig:tree} for a terminal linked-list with $7$ nodes. Each non-terminal node in the skiplist stores a key and pointers to the next and bottom nodes. The children of a node have keys $\leq$ to its key. This invariant is used to maintain order among keys. The skiplist shown in figure~\ref{fig:tree}, has two levels, with level $1$ being the leaf level. Any two consecutive levels in the skiplist maintain the invariant that keys at level $l+1$ are subsets of keys at level $l$. Addition and deletion operations on the terminal linked-list cause addition or deletion of non-terminal nodes to maintain the $O(logn)$ number of links to reach any node in the linked-list. These are referred to as re-balancing operations in this paper. They also include operations that adjust the skiplist height~(change in the number of levels).

Addition~(Insertion) operation traverses the skiplist recursively until it finds a level $1$ node with key $\leq$ the new key. Nodes on the path from the root to the leaf node are checked for 1-2-3-4 criterion and re-balancing adds a new node if a node has $5$ children. This operation avoids the creation of nodes with $\geq6$ children post addition. These are local operations which require locking nodes in $L$ shape, i.e a node and its children. The maximum number of nodes locked by a thread at any time during addition is at most $6$. Addition maintains sorted order in the terminal linked-list and new nodes are inserted after checking for duplicates. Algorithms for addition of keys are provided in algorithms~\ref{alg:alg1} and \ref{alg:alg3}. The key of the root node~(\emph{head}) is the maximum key~($2^{64}-1$). All terminal and non-terminal linked-lists are terminated with sentinel \emph{tail} nodes. If re-balancing added a node as the next neighbor of the root, then, the skiplist's height should be increased, using algorithm~\ref{alg:alg4}. Also, terminal linked-list nodes have sentinel bottom nodes. Sentinel nodes point to themselves, thereby avoiding null pointers and segmentation faults in the implementation.
We used wide unsigned integers (128-bits) to store the key and next pointer in the same data type. The 64-bit key is stored in the upper half and the 64-bit pointer is stored in the lower half of the wide integer. We used the wide unsigned integer implementation from Boost library~\cite{boost} for this purpose. Boost supports atomic operations on wide integers. Key and pointer fields were extracted from wide integers using bit masks. Nodes also contain one additional bit field : \emph{mark}. The \emph{mark} bit is set when a node is deleted and is no longer in the skiplist. Since key and next pointers are updated atomically during addition and deletion, our \emph{Find} implementation can be lock-free. In our \emph{Find} implementation, we try to obtain a set of valid nodes in $L$ shape and make traversal decisions based on their keys. If a node or its children change state, it will be detected at the next recursive step. If a node is deleted, it can be detected by reading its \emph{mark} field. If the key of a node was decreased by a re-balancing operation, the search can continue to the right, by following its next pointer.  

Deletion operation deletes a key from the terminal linked-list. Removal of nodes with deleted keys from non-terminal linked-lists is performed lazily to reduce work per deletion. If a non-terminal node~(except root) in the deletion path has exactly $2$ children, then, the number of children is increased by re-balancing, to maintain the 1-2-3-4 criteria post deletion. Deletion may perform one of two re-balancing operations \emph{borrow} or \emph{merge} at non-terminal nodes along the path from the root to the terminal linked-list. The \emph{merge} operation removes nodes from non-terminal linked-lists by merging the children of two neighboring nodes and removing one of them. In the implementation described in this paper, \emph{merge} removes the node with the higher key. The \emph{borrow} operation, borrows a node from a neighbor keeping the number of nodes constant. We implemented \emph{borrow} as a post operation of \emph{merge} to maintain the correctness of \emph{Find}. Algorithms used by deletion are provided in algorithms~\ref{alg:alg8} and \ref{alg:alg9}. We haven't provided the deletion algorithm because its control flow is similar to the addition algorithm, except that it locks a pair of adjacent nodes and their children following $LL$ pattern at every level. The pair of nodes for the next level of recursion is chosen from the union of child nodes. The previous neighbor of the child node containing the key to be deleted is the preferred partner. If a child node does not have a previous neighbor, then, its next neighbor is chosen.  Deletion of nodes can lead to a skiplist where the root node has a $0$ length interval. This is detected and the skiplist's height is decreased, using the \emph{DecreaseDepth} algorithm provided in algorithm~\ref{alg:alg9}. All function implementations \emph{Addition},~\emph{Find} and \emph{Deletion} have retry semantics i.e if a function exited prematurely due to one of the following conditions, then it retries :

\begin{enumerate}
\item Marked node.
\item The next neighbor of root is not tail.
\item Root node has $0$ length interval.
\end{enumerate}		

The \emph{DropKey} function removes a node with matching key from the terminal linked-list. The \emph{AddNode} function adds a new node to the terminal linked-list if duplicates do not exist. The \emph{CheckNodeKey} function is used to update a node's key if the keys of its children are $<$ than its key. This happens if the child node with the highest key was removed by \emph{Deletion} or \emph{merge}. The \emph{Acquire} function locks a node and \emph{Release} unlocks it. Since these are simple functions, pseudo-codes for them are not provided here. In our implementation we used a memory manager that recycles deleted nodes. We used reference counters in every node to avoid the ABA problem~\cite{herlihy}. Reference counters were incremented during every recycling operation performed on a node. 

\begin{algorithm}
\DontPrintSemicolon
\small
\KwIn {Node $n$, Key $key$}
\KwOut {Integer}
\SetKwBlock{Begin}{procedure}{end procedure}
\SetKwFunction{Acquire}{Acquire}
\SetKwFunction{isMarked}{isMarked}
\SetKwFunction{isSentinel}{isSentinel}
\SetKwFunction{Release}{Release}
\SetKwFunction{isHead}{isHead}
\SetKwFunction{AcquireChildren}{AcquireChildren}
\SetKwFunction{ReleaseChildren}{ReleaseChildren}
\SetKwFunction{CheckNodeKey}{CheckNodeKey}
\SetKwFunction{AdditionRebalance}{AdditionRebalance}
\SetKwFunction{Addition}{Addition}
\SetKwFunction{AddNode}{AddNode}
\SetKwFunction{isLeaf}{isLeaf}
\Begin($\text{Addition}{(}n,key{)}$)
{
\Acquire{n}\;
$kn\gets n.get\_keynext()$\;
nkey$\gets$kn[127:64]\;
nnext$\gets$kn[63:0]\;
\If {\isMarked{n}}
{
\Release{n}\;
\KwRet{RETRY}\;
}
\If {\isSentinel{n}}
{
\Release{n}\;
\KwRet{FALSE}\;
}
\If {\isHead{n} $\land \neg$ \isSentinel{nnext}}
{
\Release{n}\;
\KwRet{RETRY}\;
}
cnodes$\leftarrow$ null\;
b $\leftarrow$ n.bottom\;
cnodes$\leftarrow$ \AcquireChildren{n}\;
\CheckNodeKey{n,cnodes}\;

\If {nkey$<$key}
{
r$\gets$ nnext\;
\ReleaseChildren{cnodes}\;
\Release{n}\;
\KwRet{\Addition{r,key}}\;
}

\AdditionRebalance{n}\;
\If {\isLeaf{n}}
{
ret$\gets$ \AddNode{n,key}\;
\ReleaseChildren{cnodes}\;
\Release{n}\;
\KwRet{ret}\;
}
nn $\gets$ null\;
\For {d $\in$ cnodes}
{
$dkn\gets d.get\_keynext()$\;	
dkey$\gets$dkn[127:64]\;
\If {key$\leq$ dkey}
{
nn $\gets$ d\;
\bf{break}\;
}
}
\ReleaseChildren{cnodes}\;
\Release{n}\;
\KwRet{\Addition{nn,key}}\;
\caption{Skiplist : Addition\label{alg:alg1}}
}
\end{algorithm}

\begin{algorithm}
\small
\DontPrintSemicolon
\KwIn{Node $n$}	
\KwOut{void}
\SetKwBlock{Begin}{procedure}{end procedure}
\SetKwFunction{Children}{Children}
\SetKwFunction{Count}{Count}
\SetKwFunction{NewNode}{NewNode}
\Begin($\text{AdditionRebalance} {(} n {)}$)
{
cnodes $\leftarrow$ \Children{n}\;
nnodes $\leftarrow$ \Count{cnodes}\;
\If {nnodes $>$ 4}
{
nn $\gets$ \NewNode{}\;
$kn\gets n.get\_keynext()$\;
nkey$\gets$ kn[127:64]\;
nnext$\gets$kn[63:0]\;
$nn.update\_keynext(nkey,nnext)$\;
nn.bottom $\gets$ cnodes[2]\;
$kn\gets cnodes[1].get\_keynext()$\;
nkey$\gets$kn[127:64]\;
$n.update\_keynext(nkey,nn)$ \;
}
\caption{Skiplist : Add Non-Terminal Node\label{alg:alg3}}
}
\end{algorithm}

\begin{algorithm}
\small
\DontPrintSemicolon
\KwIn{}
\KwOut{}
\SetKwBlock{Begin}{procedure}{end procedure}
\SetKwFunction{Acquire}{Acquire}
\SetKwFunction{Release}{Release}
\SetKwFunction{isSentinel}{isSentinel}
\SetKwFunction{NewNode}{NewNode}
\Begin($\text{IncreaseDepth}{(}{)}$)
{
n$\gets$ head\;
\Acquire{n}\;
$kn\gets n.get\_keynext()$\;
nkey$\gets$kn[127:64]\;
nnext$\gets$kn[63:0]\;
nn$\gets$ nnext\;
\If {\isSentinel{nn}}
{
\Release{n}\;
\KwRet{}\;
}
d$\gets$ \NewNode{}\;
d.bottom$\gets$ n.bottom\;
$d.update\_keynext(nkey,nnext)$\;
n.bottom$\gets$ d\;
$n.update\_keynext(maxValue,tail)$\;
\Release{n}\;
\KwRet{}\;
}
\caption{Skiplist : Increase Depth\label{alg:alg4}}
\end{algorithm}

\begin{algorithm}
\small
\DontPrintSemicolon
\KwIn{Node $n$,Key $key$}
\KwOut{Integer}
\SetKwBlock{Begin}{procedure}{end procedure}
\SetKwFunction{isHead}{isHead}
\SetKwFunction{isSentinel}{isSentinel}
\SetKwFunction{Find}{Find}
\SetKwFunction{Count}{Count}
\Begin(\text{Find}{(}n,key{)})
{

\If {\isMarked{n}}
{
\KwRet{RETRY}\;
}

d$\gets$ n.bottom, cnodes$\gets$ null\;
keys$\gets$ null,nn$\gets$ null\;
$kn\gets n.get\_keynext()$\;
nkey $\gets$ kn[127:64],nnext $\gets$ kn[63:0]\;
	
\If {\isHead{n} $\land \neg$ \isSentinel{nnext}}
{
\KwRet{RETRY}\;
}
\If {\isSentinel{n}}
{
\KwRet{FALSE}\;
}

\If {$\neg$ \isMarked{n} $\land$ \isSentinel{d}}
{
\If {nkey=key}
{
\KwRet{TRUE}\;
}
\Else
{
\If {nkey$>$key}
{
\KwRet{FALSE}\;
}
}
}
\If {nkey $<$ key}
{
\KwRet{\Find{nnext,key}}
}

\While {true}
{
$dn\gets d.get\_keynext()$\;
\If {\isMarked{n} $\lor$ \isMarked{d}}
{
\KwRet{RETRY}\;
}
\If {dn[127:64] $\leq$ nkey $\land \neg$ \isSentinel{d}}
{
$cnodes.add(d),keys.add(dn[127:64])$
}
\If {dn[127:64] $\geq$ nkey $\lor$ \isSentinel{d}}
{
\bf{break}\;
}
d$\gets$ dn[63:0]\;
}

\If{\isHead{n}}
{
nnodes$\gets$ \Count{cnodes}\;
\If{keys[nnodes-1]$\neq$ maxkey}
{
\KwRet{RETRY}\;
}
}
c$\gets$ 0\;
\For {k$\in$ keys}
{
\If {key$\leq$ k}
{
nn$\gets$ cnodes[c]\;
\bf{break}\;
}
c$\gets$ c+1\;
}
\If{nn$\neq$ null}
{
\KwRet{\Find{nn,key}}\;
}
\Else
{
\KwRet{FALSE}\;
}
\caption{Skiplist : Algorithm for lock-free Find\label{alg:alg5}}
}
\end{algorithm}

\begin{algorithm}
\small	
\DontPrintSemicolon
\KwIn{Node $n1$,Node $n2$,Key $key$}
\KwOut{Integer}
\SetKwBlock{Begin}{procedure}{end procedure}
\SetKwFunction{Children}{Children}
\SetKwFunction{Count}{Count}
\SetKwFunction{MarkNode}{MarkNode}
\SetKwFunction{Acquire}{Acquire}
\SetKwFunction{Release}{Release}
\SetKwFunction{NewNode}{NewNode}
\Begin(\text{MergeBorrowNode}{(}n1,n2,key{)})
{
n1nodes$\gets$ \Children{n1}\;
n2nodes$\gets$ \Children{n2}\;
$n1kn \gets n1.get\_keynext()$\;
n1key $\gets$ n1kn[127:64]\;
fn $\gets$ key $\leq$ n1key ? true : false\;
b $\gets$ fn $\land$ \Count{n1nodes}=2\;
c $\gets$ $\neg$ fn $\land$ \Count{n2nodes}=2\;
\If {b $\lor$ c}
{
$n2kn \gets n2.get\_keynext()$\;
n2k $\gets$ n2kn[127:64]\;
n2n $\gets$ n2kn[63:0]\;
$n1.update\_keynext(n2k,n2n)$\;
\MarkNode{n2}\;
\Release{n2}\;
n2$\gets$ null\;
\If{b $\land$ \Count{n2nodes}$>$2}
{
nn$\gets$ \NewNode{}\;
$nn.update\_keynext(n2k,n2n)$\;
$nn.update\_bottom(n2nodes[1])$\;
$n2kn \gets n2nodes[0].get\_keynext()$\;
$n1.update\_keynext(n2kn[127:64],nn)$\;
\Acquire{nn}\;
n2$\gets$ nn\;
}
\Else
{
\If {c $\land$ \Count{n1nodes}$>$2}
{
nn$\gets$ \NewNode{}\;
$nn.update\_keynext(n2k,n2n)$\;
p $\gets$ \Count{n1nodes}\;
p1 $\gets$ p-1,p2$\gets$ p-2\;
$nn.update\_bottom(n1nodes[p1])$\;
$n2kn \gets n1nodes[p2].get\_keynext()$\;
$n1.update\_keynext(n2kn[127:64],nn)$\;
\Acquire{nn}\;
n2$\gets$ nn\;
}
}
}
\caption{Skiplist : MergeBorrow Node\label{alg:alg8}}
}
\end{algorithm}

\begin{algorithm}
\small
\DontPrintSemicolon
\KwIn{}
\KwOut{}
\SetKwBlock{Begin}{procedure}{end procedure}
\SetKwFunction{Acquire}{Acquire}
\SetKwFunction{Release}{Release}
\SetKwFunction{isSentinel}{isSentinel}
\SetKwFunction{MarkNode}{MarkNode}
\Begin(\text{DecreaseDepth}{(}{)})
{
n$\gets$ head\;
\Acquire{n}\;
$kn\gets n.get\_keynext()$\;
nkey$\gets$ kn[127:64]\;
b$\gets$ n.bottom\;
\Acquire{b}\;
$kn\gets b.get\_keynext()$\;
bkey $\gets$ kn[127:64]\;
\If {nkey=bkey$\land \neg$ \isSentinel{b}}
{
bb $\gets$ b.bottom\;
\If {bb$\neq$ b}
{
\Acquire{bb}\;
n.bottom$\gets$ bb\; 
\MarkNode{b}\;
\Release{bb}\;
}
}
\Release{b}\;
\Release{n}
}
\caption{Skiplist : Decrease Depth\label{alg:alg9}}
\end{algorithm}

All operations on this concurrent skiplist are linearizable. The linearization points in the algorithms are the statements where key and next pointer are updated, i.e a new node is added or an existing node is deleted. 
The space requirement for this data structure is $O(n)$ nodes for the terminal linked-list and at most $\sum_{i=1}^{logn}\frac{n}{2^i}$ nodes for the non-terminal linked-lists. Other data structures that can be used in place of skiplists for operations such as point location, range search and  ranking are balanced binary search trees~(BST) such as red-black trees or AVL trees and their variations. BSTs store keys and data in internal and external nodes, which makes their cost per search operation at most $log(n)$. Compared to binary trees, a 1-2-3-4 skiplist has higher concurrency due to its branching factor. Skiplist operations can be performed concurrently at every level by locking a few subset of nodes. However, search costs in skiplists implementation are never less than $log(n)$.  
Deterministic skiplists are similar in design to multi-way trees or $(a,b)$ trees, $a\geq 2$ and $b \geq 2*a$~\cite{kurt}. $(a,b)$ trees have equivalent re-balancing operations such as node splitting, node sharing and node fusing. $(a,b)$ tree implementations usually perform re-balancing bottom-up. The total number of re-balancing operations can be bounded using the bottom-up analysis described in ~\cite{kurt}.   

Define $c = min(min(2a-1,\lceil\frac{b+1}{2}\rceil)-a,b-max(2a-1,\lfloor\frac{b+1}{2}\rfloor))$. $c \geq 1$ and $b\geq 2*a$. Also define a balance function $b(v)$ on the nodes $v$ of the $(a,b)$ tree as :
\begin{equation}
b(v) = min(\rho(r)-a,b-\rho(v),c)
\end{equation}	
where $\rho(v)$ is the arity or the number of children of node $v$ and $r$ is the root node.
For the root node, $b(r)$ is defined as $min(\rho(r)-2,b-\rho(r),c)$.
The balance of a sub-tree $T$ of height $h$~($b_h(T)$) is the sum of the balance values of the nodes of $T$.
Let $T_0$ be the initial tree and $T_n$ the tree obtained after $n$ operations~(additions and deletions). Tree operations such as additions and deletions decrease the balance of nodes by a constant value. Balance can be restored by re-balancing operations. Since they are performed bottom-up, re-balancing operations at height $h$ increase the balance at $h$ and decrease the balance at $h+1$. 
Let $an_{h}$ be the number of node additions~(splitting), $mn_{h}$ the number of node merges~(fusion) and $bn_{h}$ the number of node borrows~(share) at level $h$. The recurrence for balance $b_h(T_n)$ can defined as :

\begin{equation}
\small
b_h(T_n)\geq b_h(T_0)-(an_{h-1}+mn_{h-1})+(2c*an_{h}+(c+1)*mn_{h})
\end{equation}	

Rearranging, and setting $b_h(T_0)=0$ for $h\geq1$ for empty tree $T_0$, we get,

\begin{equation}
\small
an_{h}+mn_{h} \leq \frac{an_{h-1}+mn_{h-1}}{c+1}+\frac{b_h(T_n)}{c+1}
\end{equation}

\begin{equation}
\small
bn_{h} \leq mn_{h-1}-mn_{h} 
\end{equation}

Solving the recurrences, we can bound the total number of re-balancing operations up to height $h$ in an $(a,b)$ tree to $2*(c+2)*\frac{n}{(c+1)^h}$ for $n$ additions and deletions. This value decreases exponentially with increasing value of $h$. The complete proof can be found in ~\cite{kurt}. From the analysis of $(a,b)$ trees, we observe that most re-balancing operations are performed at the lower levels of the data structure. Since the lower levels have higher concurrency, several re-balancing operations can be performed in parallel. 
In our concurrent skiplist design, re-balancing is performed proactively during top-down traversals. Alternatively, re-balancing can be performed using bottom-up traversals post skiplist operations, which is the implementation described by Munro and Sedgewick~\cite{sedgewick}. In both cases, the total number of re-balancing operations is linear in the number of additions and deletions.
The recurrences described above hold for both top-down and bottom-up re-balancing if $b>2*a+1$. When $b=2*a$ or $b=2*a+1$~(1-2-3-4 skiplist), the above recurrences are not valid for top-down re-balancing. For example, for $a=2$,$b=5$, a node with $5$ children, when split, will create nodes with $2$ and $3$ children, where the $2$ nodes will require re-balancing during deletions. But the number of re-balancing operations will still be linear in the number of operations.  

We did not use bottom-up re-balancing because it requires two traversals per operation. It will be difficult to maintain the correctness of lock-free \emph{Find} along with addition and deletion traversals in opposite directions. Multi-way search trees~\cite{abtrees} discusses a concurrent implementation which performs re-balancing bottom-up but the \emph{Find} implementation uses locks. The lock-free $B^{+}$-tree design from~\cite{lockfreebtree} maintains correctness by deleting nodes after every re-balancing operation.

\section{Unbounded Lock-free Queues\label{queue}}

We chose lock-free queues for this discussion because they are useful for load balancing workloads within and across nodes in many-core processors. Boost provides a lock-free queue implementation that allocates memory in blocks. Their implementation is based on the lock-free algorithm from \cite{MichealScott}. Although \emph{push} and \emph{pop} are lock-free, memory management operations such as allocation of additional blocks, uses coarse locks on the queue. 
Some of the reasons for poor performance of concurrent queues on many-core processors are the the following : 

\begin{enumerate}
\item Use of coarse locks on the queue 
\item Frequent dynamic memory allocation 
\item Cache misses, especially in linked-list based implementations 
\end{enumerate}


\begin{algorithm}
\DontPrintSemicolon
\small
\KwIn {Integer}
\KwOut {Boolean}
\SetKwBlock{Begin}{procedure}{end procedure}
\SetKwFunction{atomicAdd}{atomicAdd}
\SetKwFunction{atomicCAS}{atomicCAS}
\SetKwFunction{addNode}{addNode}
\Begin($\text{push}{(}r{)}$)
{
n $\leftarrow$ cn\;
b $\leftarrow$ False\;
\If {$\neg$list[n].wclosed}
{
p $\leftarrow$ \atomicAdd{list[n].rear,1}\;
\If{p $<$ list[n].size}
{
  pv $\leftarrow$ list[n].fe[p]\;
\If{pv$=$0}
{
   list[n].data $\leftarrow$ r\;
   pf $\leftarrow$ \atomicAdd{list[n].fe[p],1}\;
   \If{pf$=$0}
   {
     b $\leftarrow$ True\;
   }
}
}
\Else
{
  pv $\leftarrow$ \atomicCAS{list[n].wclosed,0,1}\;
}
}
\Else
{
nn $\leftarrow$ list[n].next\;
\If{nn $neq$ INTMAX}
{
 pv $\leftarrow$ \atomicCAS{cn,INTMAX,nn}\;
}
\Else
{
\addNode{}\;
}
}
\KwRet{b}\;

\caption{LockfreeQueue : Push\label{alg:alg10}}
}
\end{algorithm}

\begin{algorithm}
\small
\DontPrintSemicolon
\KwIn{Node $n$}
\KwOut{void}
\SetKwBlock{Begin}{procedure}{end procedure}
\SetKwFunction{Count}{Count}
\SetKwFunction{NewNode}{NewNode}
\SetKwFunction{atomicAdd}{atomicAdd}
\SetKwFunction{atomicCAS}{atomicCAS}
\SetKwFunction{break}{break}
\Begin($\text{addNode} {(}{)}$)
{
n $\leftarrow$ cn\;
bn $\leftarrow$ list[n].next\;
b $\leftarrow$ False\;
\If{bn$=$INTMAX $\wedge$list[n].rear $\geq$ list[n].size}
{
 e $\leftarrow$ -1\;
\For{i $\in$ maxnodes}
{
\If{use[i]$=$0}
{
p$\leftarrow$ \atomicCAS{use[i],0,1}\;
\If{p$=$0}
{
   e$\leftarrow$ i\;
   \break\;
}
}
}
\If{e$=$-1}
{\KwRet{b}\;}

p $\leftarrow$ \atomicCAS{list[n].next,INTMAX,e}\;
\If{p $\neq$ INTMAX}
{
pn $\leftarrow$ \atomicCAS{use[e],1,0}\;
}
\Else
{
  b $\leftarrow$ True\;
}
}
\KwRet{b}\;
\caption{LockFree Queue : Add Node\label{alg:alg11}}
}
\end{algorithm}

\begin{algorithm}
\small
\DontPrintSemicolon
\KwIn{}
\KwOut{}
\SetKwBlock{Begin}{procedure}{end procedure}
\SetKwFunction{atomicAdd}{atomicAdd}
\SetKwFunction{atomicCAS}{atomicCAS}
\SetKwFunction{deleteNode}{deleteNode}
\Begin($\text{Pop}{(}v{)}$)
{
n $\leftarrow$ listhead\;
b $\leftarrow$ False\;
\If{$\neg$list[n].rclosed}
{
  f $\leftarrow$ list[n].front\;
  r $\leftarrow$ list[n].rear\;
\If{f $<$ list[n].size}
{
\If{f $<$ r}
{
 p $\leftarrow$ \atomicAdd{list[n].front,1}\;
\If{p $<$list[n].rear $\wedge$ p $<$ list[n].size}
{
fe $\leftarrow$ list[n].fe[p]\;
nt $\leftarrow$ 2\;
pv $\leftarrow$ \atomicCAS{list[n].fe[p],fe,nt}\;
        \If{pv$=$1}
        {
           v $\leftarrow$ list[n].data[p]\;
           b $\leftarrow$ True\;
        }
}
\Else
{
\If{p $<$ list[n].size}
{
fe $\leftarrow$ list[n].fe[p]\;
pv $\leftarrow$ -1\;

\While{pv $\neq$ fe}
{
fe $\leftarrow$ list[n].fe[p]\;
nt $\leftarrow$ 2\;
pv $\leftarrow$ \atomicCAS{list[n].fe[p],fe,nt}\;
}

\If{pv $=$ 1}
{
  v $\leftarrow$ list[n].data[p]\;
  b $\leftarrow$ True\;
}
}
}
}
}
\Else
{
p $\leftarrow$ \atomicCAS{list[n].rclosed,0,1}\;
}
}
\Else
{
\If{list[n].wclosed$=$1 $\wedge$ list[n].rclosed$=$1}
{
\deleteNode{}\;
}
}
\KwRet{b}\;
}
\caption{LockFree Queue : Pop\label{alg:alg12}}
\end{algorithm}

\begin{algorithm}
\small
\DontPrintSemicolon
\KwIn{}
\KwOut{Integer}
\SetKwBlock{Begin}{procedure}{end procedure}
\SetKwFunction{atomicAdd}{atomicAdd}
\SetKwFunction{atomicCAS}{atomicCAS}
\Begin(\text{deleteNode}{(}{)})
{
 n $\leftarrow$ listhead\;
 b $\leftarrow$ False\;
\If{list[n].rclosed$=$1 $\wedge$ list[n].wclosed$=$1 $\wedge$ n$\neq$cn}
{
  pn $\leftarrow$ list[n].next\;
\If{pn $\neq$ INTMAX}
{
p $\leftarrow$ \atomicCAS{listhead,n,pn}\;
\If{p$=$n}
{
list[n].front $\leftarrow$ 0\;
list[n].rear $\leftarrow$ 0\;
list[n].rclosed $\leftarrow$ 0\;
list[n].wclosed $\leftarrow$ 0\;
\For{i $\in$ blksize}
{
  list[n].fe[i] $\leftarrow$ 0\;
}
list[n].next $\leftarrow$ INTMAX\;
nn $\leftarrow$ \atomicCAS{use[n],1,0}\;
}
}
}

\caption{Lockfree Queue : Delete Node\label{alg:alg13}}
}
\end{algorithm}

Linked-list based lock-free queue algorithms perform more computation per \emph{push} or \emph{pop}, because of pointer updates using compare-and-swap~(CAS) instructions which have to be retried on failure. Depending on the algorithm each queue operation may require two CAS operations~(one for updating next pointer and the other for updating head/tail~(front/rear) values). Besides CAS retries, cache misses and page faults affect the performance of this implementation because of the lack of spatial locality. Spatial locality can be improved by allocating memory in blocks and by recycling used blocks. Since this reduces the total memory allocated, it improves the cache behavior of the program.  

We designed a lockfree queue which used arrays instead of linked-lists, because \emph{front} and \emph{rear} pointers can be replaced by integers whose values are updated using fetch-add instructions. This allows the non-blocking progress of multiple \emph{Push} and \emph{Pop} instructions which improves concurrency. The queue arrays were allocated in blocks and recycled. Our design is based on Linearizable Non-blocking FIFO queues~(LCRQ)~\cite{lcrq} with memory management. \emph{Front} and \emph{Rear} values are incremented monotonically during \emph{push} and \emph{pop} operations. In our implementation we created a memory pool by pre-allocating a set of blocks. New blocks were introduced into the queue from the memory pool whenever it was full and empty blocks at the head of the queue were deleted and returned to the memory pool. \emph{Push} and \emph{Pop} methods are described in the pseudocodes~\ref{alg:alg10} and \ref{alg:alg12}, while \emph{AddNode} and \emph{DeleteNode} are described in algorithms~\ref{alg:alg11} and \ref{alg:alg13}. The queue is comprised of a vector of blocks termed $list$, where each block has $front$, $rear$, $size$, $next$, $use$, $data$, $fe$, $rclosed$ and $wclosed$ integer fields. The $next$ field stores the id of the next active block in the queue. The $data$ and $fe$ fields are used for storing and retrieving data, while the remaining fields are used for controlling accesses to the block. In addition to $front$, $rear$ and $size$, we also have $rclosed$ and $wclosed$ to indicate the completion of block use. $cn$ and $listhead$ store the ids of the most recent and least recent active blocks in the queue. $maxnodes$ is the maximum number of pre-allocated blocks and $blksize$ is the size of a block. 

Unlike linked-list based algorithms, the completion of \emph{push} and \emph{pop} operations are not well-defined in array based algorithms. In this implementation we used Full/Empty~($fe$) arrays to signal completion of read/write operations to the $data$ field. Another option is to use wide data~($> 64$ bits) with separate $data$ and $fe$ bit fields, but at the cost of reduced number of available bits for storing data. Since \emph{rear}/\emph{front} are integer values, it was necessary to ensure that a successful \emph{pop} operation read valid data and that \emph{pop} operations do not miss stored data due to instances when $front$ gets ahead of $rear$. The values of $fe$ fields were used to exchange the signals necessary for validating \emph{push}es and \emph{pop}s. Let $N=n_1+n_2$, where $n_1$ and $n_2$ are the number of \emph{push} and \emph{pop} operations. Let $C$ be the size of a block. The number of blocks in use is at most $\lceil\frac{n_1}{C}\rceil$ and at least $\max(1,\lceil\frac{n_1-n_2}{C}\rceil)$. The maximum number of blocks in use depends on the sequence of \emph{push} and \emph{pop} operations. This analysis holds for both sequential and concurrent queues because an execution of a concurrent queue can be linearized into a sequence of queue operations that match a sequential execution where the linearization points are the instructions that update \emph{front} and \emph{rear} values.

\begin{figure}[!tbph]
\begin{center}
\includegraphics[width=2.8in,height=1.8in]{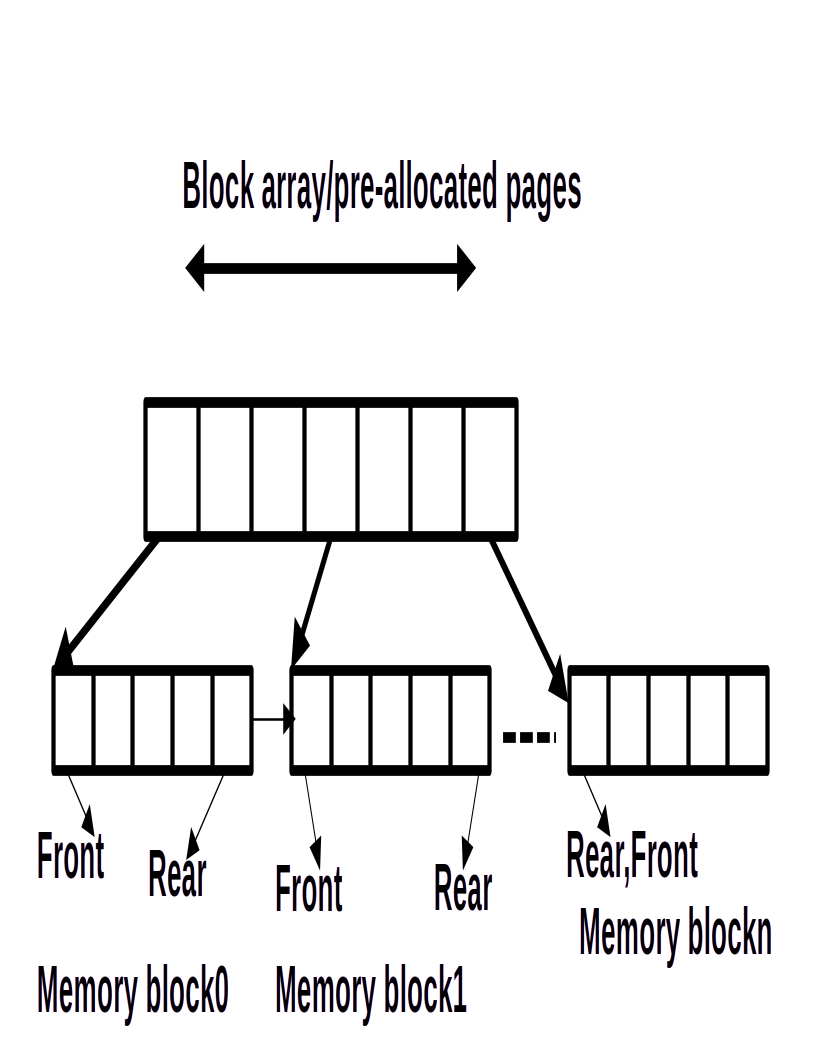}
\caption{Lock-free Queue : Memory Allocation\label{fig:queue}}
\end{center}
\end{figure}

The lock-free queue is shown in figure~\ref{fig:queue}.

\section{Experiments : Unbounded Lock-free Queues \label{queueperf}}

\begin{figure}[!tbph]
\begin{center}
\includegraphics[width=3.4in,height=2.5in]{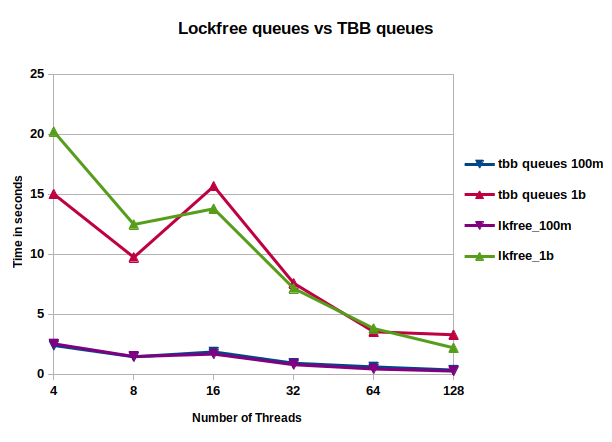}
\caption{Lock-free Queues : Performance for 100m and 1b operations on AMD Milan\label{fig:queueres}}
\end{center}
\end{figure}

We compared the performance of Boost and TBB lock-free queues with our implementation. The queues were tested with different workloads consisting of $100m$ and $1b$ operations on integers. The workloads had approximately $50\%$ \emph{push} and \emph{pop} operations. We used a vector of queues, one per thread, for testing these workloads. We pinned threads to CPUs in the order of thread ids, e.g thread 0 was pinned to CPU0. An AMD Milan CPU node had 8 NUMA nodes each associated with 16 CPUs, i.e CPUs 0-15 belonged to NUMA node 0, 16-31 belong to NUMA node 1 and so on. The Fujitsu A64FX CPU has 4 NUMA nodes with 12 cores each. During memory allocation, we used huge pages and the scalable memory allocator from TBB for allocating blocks. Threads chose random queues within their NUMA regions for \emph{Push}es and their local queues for \emph{Pop}s. Therefore, all memory accesses were limited to their NUMA nodes. Our implementation showed strong scaling and performed better than the boost implementation for all workloads. The results from Boost queues are not presented here for brevity. The block size was fixed at $8192$ integers for all experiments. The results presented in table~\ref{tab:qtable1} and graph~\ref{fig:queueres} compare the performance of TBB concurrent queues and our implementation~\ref{fig:queueres}, referred to as \emph{lkfree}. All observations are averaged over $5$ repetitions. The TBB implementation is also based on the LCRQ algorithm~\cite{lcrq}. To describe briefly, the LCRQ algorithm uses a linked-list of circular buffers~(micro-queues), with \emph{front} and \emph{rear} values managed using fetch-add instructions. However, the LCRQ algorithm does not manage memory. The differences in performance between TBB and our implementation are due to the additional work performed by our algorithm for recycling blocks. Our implementation retries queue operations in the following scenarios :

\begin{enumerate}
\item A \emph{Push} finds queue full, allocates and links new block and retries.
\item A \emph{Pop} unlinks and recycles empty block.
\item A \emph{Push} finds \emph{front} is ahead of its \emph{rear} value
\item A \emph{Pop} finds a data field has invalid data~(incomplete \emph{Push} or \emph{front} is ahead of \emph{rear})
\end{enumerate}		

\begin{table}[!tbph]
\scriptsize
\begin{center}
\begin{tabular}{|c|c|c|c|c|}
\hline
\textbf{\#threads}&\textbf{100mtbb(s)}&\textbf{100mlkfree(s)}&\textbf{1btbb(s)}&\textbf{1blkfree(s)}\\
\hline
4 & 2.525576 &3.23806& 14.9945 & 20.19996\\
\hline
8 &1.468532&2.033946&9.728728 &12.46478\\
\hline
16 &1.672976 &2.378378 &15.65188 &13.7761 \\
\hline
32 &0.7895414 &1.286334 &7.565792 & 7.139884\\
\hline
64 &0.4291294 &0.6874498 &3.532416 &3.800926 \\
\hline
128 &0.2574812 &0.3819218 &3.279696&2.18968\\
\hline
\end{tabular}
\vspace{1mm}
\caption{Concurrent Queue Performance for 100million and 1billion workloads on AMD Milan\label{tab:qtable1}}
\end{center}
\end{table}	

\begin{table}[!tbph]
\scriptsize
\begin{center}
\begin{tabular}{|c|c|c|}
\hline
\textbf{\#threads}&\textbf{10mlkfree(s)}&\textbf{1blkfree(s)}\\
\hline
4 & 0.7984978 & 78.60762 \\
\hline
8 &0.3012404 & 28.55252 \\
\hline
16 & 0.2158552 & 20.4112 \\
\hline
32 & 0.1170594 & 10.59888 \\
\hline
48 & 0.0830941 & 7.149244\\
\hline
\end{tabular}
\vspace{1mm}
\caption{Concurrent Queue Performance for 10million and 1billion workloads on Fujitsu A64FX\label{tab:qtable1fu}}
\end{center}
\end{table}

\begin{figure}[!tbph]
\begin{center}
\includegraphics[width=2.8in,height=2in]{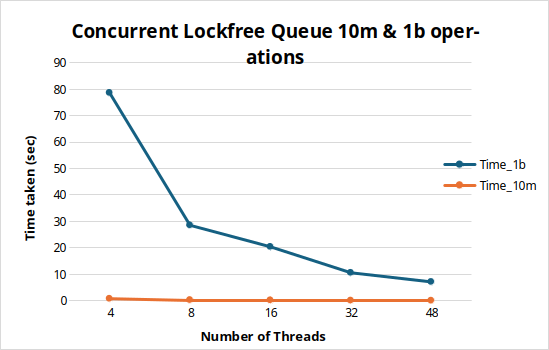}
\caption{Lock-free Queues : Performance for 10m and 1b operations on Fujitsu A64FX\label{fig:queueresfu}}
\end{center}
\end{figure}

The performance of the concurrent lockfree queue on Fujitsu A64FX is tabulated in table~\ref{tab:qtable1fu} and plotted in the graph in figure~\ref{fig:queueresfu}.
Our algorithm had good strong scaling performance on both processors. The LCRQ algorithm and memory recycling are carefully designed to scale well on many-core numa processors. From the results on both processors, we conclude that lower contention, improved cache locality and reduced remote numa accesses are essential for the strong scaling of shared memory programs on many-core NUMA processors. 

\section{Memory Management\label{mem}}

In this section we discuss the memory management module used in our skiplist and hash table implementations. In multi-threaded programs memory allocation causes overheads when a large number of threads make concurrent calls to \emph{malloc} and \emph{free}. This can affect the scalability of such programs. A memory manager that reduces the total memory used by a concurrent program and performs recycling can improve scalability by improving average spatial and temporal localities in the program. TBB has a scalable memory allocator~\cite{TBB} which is common to all programs and modules. Besides using a scalable allocator, it is useful if data structures manage their own memory for better page and cache locality within each module.  In the programs described in this article we used a concurrent memory manager which reduced the total number of calls to \emph{malloc} by allocating memory in blocks. Deletion of allocated blocks is performed when the data structure goes out of scope. When a node is deleted in the hash table or skiplist, its pointer is enqueued to a concurrent lock-free queue. Work-sharing and recycling are the two strategies used for memory management. Suppose a program requires $N$ entities. Requests for new entities are made using \emph{new} and requests for removal of entities are made using \emph{delete}. All \emph{new}s are matched by \emph{delete}s. Let $C$ be the size of a block. The number of blocks allocated is at most $\lceil\frac{N}{C}\rceil$ and at least $1$. Since entities are recycled, the number of blocks allocated depends on the order of \emph{new}s and \emph{delete}s in the program. The maximum number of blocks are allocated when all \emph{new}s precede \emph{delete}s. The number of blocks allocated is $1$ when \emph{new} and \emph{delete} alternate. We can consider \emph{new}s and \emph{delete}s as two different sequences of lengths at most $N$. 
Of all such sequences, valid sequences are those where the number of \emph{delete}s are less than or equal to the number of \emph{new}s. The average number of blocks in use by this memory manager is given by the equation below, where $k$ is the number of \emph{new} requests and $i$ is the number of \emph{delete} requests :  

\begin{equation}
\frac{\sum_{k=1}^{N}\sum_{i=0}^{k} \lceil\frac{k-i}{C}\rceil}{\sum_{i=1}^{N}i}
\end{equation}	

Although the analysis is provided for a sequential program, it is also valid for a concurrent program. The \emph{new} and \emph{delete} calls made by a concurrent program to the memory manager are linearizable. The linearization point for \emph{new} is the atomic operation that increments the current valid address in a block or the \emph{pop} operation that supplies a recycled node from the lock-free queue. The linearization point for \emph{delete} is the \emph{push} operation to the lock-free queue. Linearization ensures the correctness of the memory manager. It guarantees that concurrent \emph{new} requests obtain unique memory locations. If multiple threads detect an \emph{empty} queue, it could lead to allocation of extra blocks. Memory management using recycling also leads to better programs that do not crash from segmentation faults. 

\section{Experiments : SkipList~(1-2-3-4 Trees) \label{skiplistperf}}

\begin{table}[!tbph]
\scriptsize
\begin{center}
\begin{tabular}{|c|c|c|}
\hline
\bf{\#threads} & \bf{10mRWlocks} & \bf{10mlkfreefind} \\
\hline
4 & 16.3483 & 13.70978\\
\hline
8 & 9.237172 & 7.842358 \\
\hline
16  & 11.7282 & 8.181222\\
\hline
32 & 6.77715 & 5.31692\\
\hline
64 & 4.614454 & 4.869106\\
\hline
128 & 4.248924 & 3.739122\\
\hline
\end{tabular}
\vspace{1mm}
\caption{Skiplist Performance 10m operations on AMD Milan\label{tab:skiplist1}}
\end{center}
\end{table}

\begin{figure}[!tbph]
\begin{center}
\includegraphics[width=3.5in,height=2.4in]{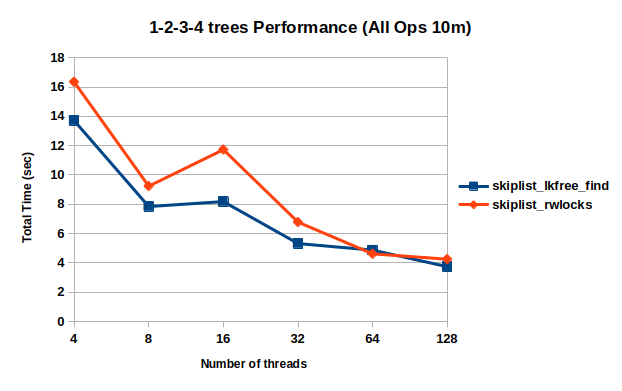}
\caption{Skiplist Performance 10m operations on AMD Milan\label{fig:skiplist1}}
\end{center}
\end{figure}

\begin{table}[!tbph]
\scriptsize
\begin{center}
\begin{tabular}{|c|c|c|c|c|}
\hline
\bf{\#threads} & \bf{RWL(IF)} & \bf{lkfreefind(IF)} & \bf{RWL(IFE)} & \bf{lkfreefind(IFE)} \\
\hline
4 & 195.069 &  138.496 & 207.9766 & 136.8524 \\
\hline
8 & 104.2194 & 75.27658 & 102.8858 & 75.15104 \\
\hline
16 & 103.9242 & 71.53346 & 101.54936 & 88.02024 \\
\hline
32 & 80.00542 & 45.49626 & 60.25536 & 56.98748 \\
\hline
64 & 54.5701 & 37.90108 &41.77146 & 47.41808 \\
\hline
128 & 40.8587 & 34.28502 & 39.33168 & 32.7872 \\
\hline
\end{tabular}
\vspace{1mm}
\caption{Skiplist Performance 100m ops~(Inserts~(I),Finds~(F),Erases~(E)) on AMD Milan\label{tab:skiplist2}}
\end{center}
\end{table}

\begin{figure}[!tbph]
\begin{center}	
\includegraphics[width=3.3in,height=2.5in]{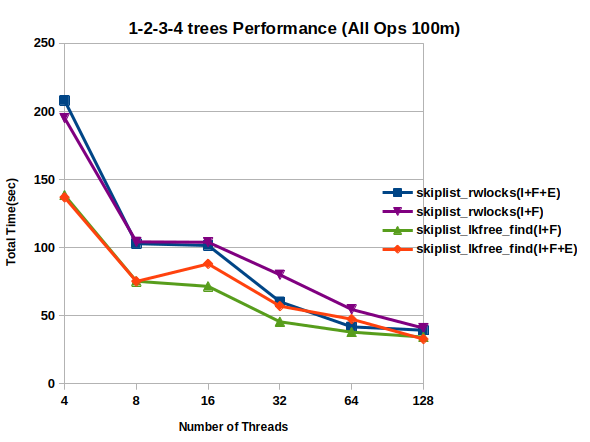}
\caption{Skiplist Performance for 100m operations~(Inserts,Finds,Erases) on AMD Milan\label{fig:skiplistall}}
\end{center}
\end{figure}

\begin{table}[!tbph]
\scriptsize
\begin{center}
\begin{tabular}{|c|c|c|}
\hline
\bf{\#threads} & \bf{lkfreefind} & \bf{lkfreeRandomSL} \\
\hline
4 & 138.496 & 43.7999 \\
\hline
8 & 75.27658 & 23.00286 \\
\hline
16 & 71.53346 & 17.16074 \\
\hline
32 & 45.49626 & 8.108614 \\
\hline
64 & 37.90108 & 4.343792 \\
\hline
128 & 34.28502 & 2.863776\\
\hline
\end{tabular}
\vspace{1mm}
\caption{Lockfree RandomSkiplists~(lkfreeRandomSL) vs Deterministic 1-2-3-4 Trees~(lkfreefind) 100m operations on AMD Milan\label{tab:randomskiplist}}
\end{center}
\end{table}

\begin{figure}[!tbph]
\begin{center}
\includegraphics[width=3.5in,height=2.4in]{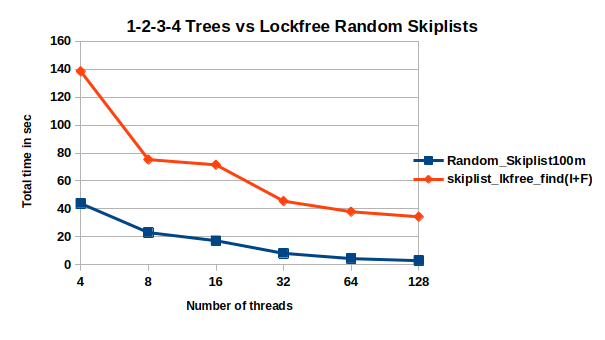}
\caption{Lockfree RandomSkiplists vs Deterministic 1-2-3-4 Trees for 100m operations on AMD Milan\label{fig:randomskiplist}}
\end{center}
\end{figure}

\begin{table}[!tbph]
\scriptsize
\begin{center}
\begin{tabular}{|c|c|}
\hline
\bf{\#threads} & \bf{lkfreeRandomSL} \\
\hline
4 & 81.47448\\
\hline
8 & 41.08576\\
\hline
16 & 20.8052 \\
\hline
32 & 10.41858\\
\hline
48 & 7.522188\\
\hline
\end{tabular}
\vspace{1mm}
\caption{Lockfree RandomSkiplists Performance for 100m operations on Fujitsu A64FX\label{tab:randomskiplistfu}}
\end{center}
\end{table}

\begin{figure}[!tbph]
\begin{center}
\includegraphics[width=3in,height=2in]{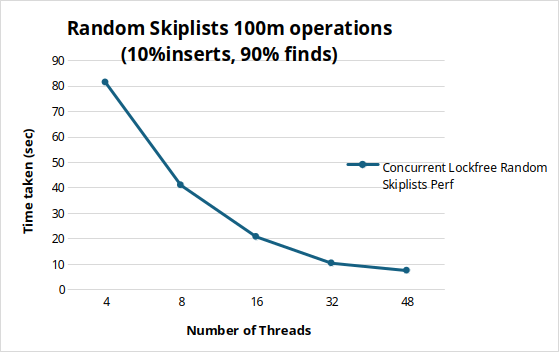}
\end{center}
\caption{Lockfree RandomSkiplist Performance for 100m operations on Fujitsu A64FX\label{fig:randomskiplistfu}}
\end{figure}

We partitioned the skiplist into one skiplist per NUMA node~(skiplists0-7) for the AMD processor and 8 skiplists per NUMA node for the Fujitsu processor. The keys used for skiplist operations were 64-bit unsigned integers generated using hash functions from boost~\cite{boost}. The key space was partitioned across skiplists using 3 MSBs~(bits 63-61) for AMD and 5 MSBs for Fujitsu. Let $n$ be the number of skiplists, $n_u$ the number of NUMA nodes in use and $n_{cpu}$, the number of CPUs per NUMA node. Let $T$ be the number of threads, $S_i$ the $i^{th}$ skiplist and $n_{s_i}$, the NUMA node it is assigned to.
\begin{equation}
n_{u} = \lceil\frac{T}{n_{cpu}}\rceil
\end{equation}
\begin{equation}
n_{s_i} = S_i\mod n_{u}
\end{equation}

We used lock-free queues, one per thread for distributing keys. The queues distributed keys with upper 3-bits~(5-bits) equal to $S_i$ to a random thread in $n_{s_i}$. For example, $T=32$, $n_{cpu}=16$, $n_u=2$, the $8$ skiplists were divided into odd-even groups on AMD. All keys for even-numbered skiplists were serviced by threads pinned to NUMA node 0, while threads pinned to NUMA node 1, accessed odd-numbered skiplists. For large skiplists, hierarchical decomposition reduced the number of pages accessed per NUMA node and lowered overheads from page faults. 

We tested the concurrent skiplist using two types of workloads :

\begin{enumerate}
\item Workload1 : 10\% insertions and 90\% find
\item Workload2 : 10\% insertions, 90\% find and 0.2\% erasures
\end{enumerate}		

The concurrent skiplists showed strong scaling for all workloads, although performance dropped at high thread counts. The graph in figure~\ref{fig:skiplist1} and the observations in table~\ref{tab:skiplist1} show the results for workload1 for 10 million operations. Figure~\ref{fig:skiplistall} and table~\ref{tab:skiplist2} are the results for 100 million skiplist operations for workloads1 \& 2. In our experiments, we compared the performance of our deterministic skiplist implementation~(\emph{lkfreefind}) with a baseline version with Read/Write locks~(\emph{RWL}). All observations reported in this sections are averaged over $5$ repetitions of each experiment. The baseline version used RW locks~(Intel TBB RW locks) for all three operations. Our deterministic skiplist with \emph{lockfree\_find} is more scalable compared to the baseline for workload1. For workload2 with 100m operations, our implementation is slower than the baseline for $64$ threads. The implementation with \emph{lockfree\_find} is affected by the overhead of retries for failed \emph{find} operations in workload2. Skiplists do not have spatial or temporal locality in their traversals. A root to terminal node path can occupy several cache lines. Following different paths leads to cache misses which affect their performance at high thread counts. Inspite of the improvement from \emph{lockfree\_find}, the performance of the concurrent deterministic skiplist is affected by contention due to locks. 
From our observations, we concluded that concurrent randomized skiplists are better than concurrent deterministic 1-2-3-4 trees for large workloads. We implemented the lockfree randomized skiplist algorithm developed by~\cite{randomskiplist} alongwith the memory manager described in section~\ref{mem}. Since randomized skiplists do not require re-balancing operations to maintain $logn$ height criterion, they perform less work per operation and they can be made lockfree. The results from our experiments on both implementations using workloads with 100 million skiplist operations are tabulated in table~\ref{tab:randomskiplist} and plotted in figure~\ref{fig:randomskiplist}. Our deterministic 1-2-3-4 tree design has high contention due to locks acquired in $L$ shape during \emph{Addition} and $LL$ shape during \emph{Deletion}. A lockfree implementation similar to lockfree B-trees~\cite{lockfreebtree} may be a good approach for obtaining a scalable deterministic tree. But such implementations are affected by the overheads of re-balancing. Therefore, randomized skiplists have higher concurrency than any balanced concurrent tree implementations and they can be used to replace binary trees in parallel applications. Our randomized skiplist implementation showed strong scaling performance on both AMD and Fujitsu processors. The performance of the randomized skiplist on A64FX is tabulated in table~\ref{tab:randomskiplistfu} and plotted in the graph in figure~\ref{fig:randomskiplistfu}. It can be used in place of balanced trees for storing and accessing sorted data concurrently on many-core NUMA processors. 
All three skiplist implementations discussed in this section used the memory management methods described in this paper. Each skiplist had its own memory manager, local to each NUMA node. We also used huge pages and the scalable \emph{tbbmalloc} in our evaluations. 

\section{Hash Tables\label{ht}}
Parallel hash tables are not preferred by many applications in computational science for storing and retrieving data due to random memory accesses, resizing and rehashing, which affects strong scaling of these applications. Some of the high performance parallel hash table implementations we found are cuckoo hashing~\cite{cuckoo}, hopscotch hashing~\cite{hopscotch} and split-order~\cite{cilk}. We chose split-order tables provided by cilk~(TBB) for comparison in this section because it is a dynamic hash table implementation with resizing and rehashing. Concurrent cuckoo and hopscotch hash table implementations do not have scalable resizing options. Our goal was to identify a flexible dynamic concurrent hash table implementation with scalability on many-core processors. 

In our initial evaluation, we observed the following reasons for poor performance of concurrent hash table implementations :

\begin{enumerate}
\item Resizing and rehashing : Most implementations of dynamic concurrent hash tables, lock the entire hash table during these operations, which affects scalability. Implementations in which rehashing is followed by data migration has overheads from page faults and cache misses.  
\item Page faults and cache misses : The performance of large concurrent hash tables suffer from page faults and cache misses caused by random memory accesses, especially on many-core nodes with multiple NUMA nodes.
\item Memory management : Large workloads which perform several insertions and deletions have overheads from several concurrent calls to mallocs and frees.
\item Collisions : Some of the data structures used for managing collisions in hash tables are lists, binary trees and multi-level hash tables. These data structures should maintain the constant cost per operation invariant for hash table operations. The auxiliary data structures should also support concurrent reads and writes.
\end{enumerate}

Using singly linked-lists or binary trees and fine-grained locks per slot reduced contention and provided basic scalable hash table implementations. But for high thread counts and large workloads, these auxiliary data structures grow in size and affect performance due to contention. Based on our observations we finalized the following concurrent multi-reader multi-writer hash tables for discussion in this section.
\begin{enumerate}
\item Fixed number of slots and binary trees per slot to resolve collisions.
\item Two-level hash tables with binary trees to resolve collisions. 
\item Split-order hash tables with singly linked-lists for collision resolution.
\item Two-level split-order hash tables with singly linked-lists for collision resolution.
\end{enumerate}

Define key-space $K_p=[0,2^m)$ for $m-bit$ keys. $K_p$ was partitioned equally across all available NUMA nodes with a concurrent hash table per node. We used a vector of lock-free queues, one per thread, for distributing keys to their appropriate NUMA nodes and load balancing the workload. For $n$ NUMA nodes, $log(n)$ bits from the MSB of the key were used to locate the NUMA node. The key was pushed to a random queue in the node. Threads popped keys from their local queues and performed operations on the nearest hash table. For scalability, we used the concurrent lock-free queues from the previous section. Like in the previous section, the number of NUMA-nodes in use depends on the number of threads. NUMA-nodes varied from $1$~(4 threads) to $8$~(128 threads).  

\section{Experiments : Hash Tables \label{htperf}}
In this section we compare the performance of three concurrent hash table implementations for three random workloads with $10m$, $100m$ and $1b$ operations with $50\%$ \emph{insert} and \emph{find} operations. We used hash functions from Boost~\cite{boost} to generate hash values by scrambling 64-bit integers. Let $k$ be a key and $H(k)$ its hash key. For hash table size $M$, the slot $s$ for $k$ is provided by the equation :

\begin{equation}
s = H(k)modM
\end{equation}

Since the bits are randomized using $H(k)$, we used power-of-two values for $M$, so that the modulo operation is the last $log(M)$ bits of $H(k)$. Let $M$ be the number of slots and $N$ the number of entries in the hash table. For all implementations discussed here, $N > M$ the mean load per slot was $\frac{N}{M}$ for the values for $M$ and $N$ used in our tests~\cite{sedgewickanalysis}. The hash function distributed values without clustering and all slots were load balanced with approximately $\frac{N}{M}$ entries.

\begin{figure}[!tbph]
\begin{center}	
\includegraphics[width=3.6in,height=2.6in]{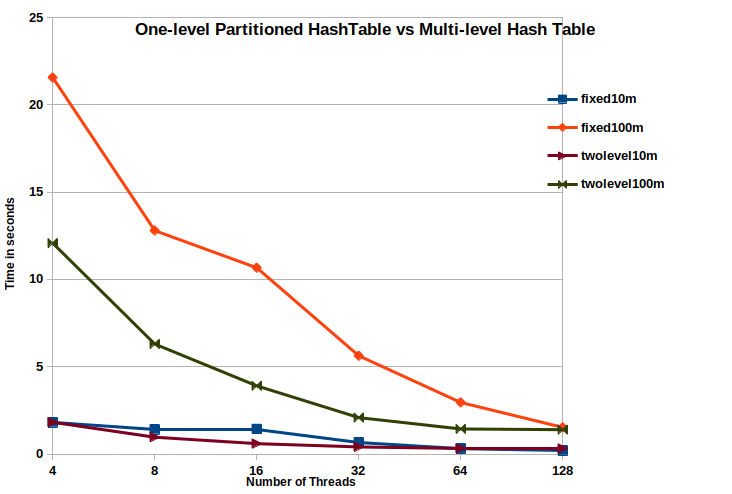}
\caption{Fixed-size Hash Tables vs Two-level Hash Tables on AMD Milan\label{fig:oneml}}
\end{center}
\end{figure}

We did not include $erase$ in this workload because it would reduce the total size of the hash table and skew the cache misses and page faults we intended to capture in our experiments. The hash table implementations are described in detail here. 
The first version has fixed number of slots and binary trees for resolving collisions. The total number of slots was kept constant and partitioned equally across the NUMA nodes in use. Although this implementation showed scalability with increasing thread counts, its execution time for large workloads was affected by large binary tree traversals per slot. The second version used another level of hash tables per slot and binary trees at the second level. Let $M_1$ be the number of slots in the hash table and $M_2$, the number of slots in the second level tables. This implementation used different subsets of bits from $H(k)$ for the two levels. The lower $log(M_1)$ bits were used for computing first level slots and the next $log(M_2)$ bits were used to locate second level slots. The two-level implementation had read-write locks per slot at both levels. During concurrent hash table operations shared read locks are acquired at first level and read-write locks at second level tables.  
The two-level hash tables had shorter binary trees and the implementation performed better than the first version at the cost of increased slots and memory allocation for slots. The graph in figure~\ref{fig:oneml} and the observations in table~\ref{tab:tableht1} show the performance of the two concurrent hash table implementations. The two-level hash tables performed better than the fixed size hash table for all workloads. We used $8192$ slots for both tables for $10m$ and $100m$ workloads. The two-level tables had either $1$ or $2048$ slots at the second level. The number of slots per NUMA node was $\frac{8192}{n}$ for $n$ NUMA nodes in both implementations. 

\begin{table}[!tbph]
\scriptsize
\begin{center}	
\begin{tabular}{|c|c|c|c|c|}
\hline
\textbf{\#threads}&\textbf{fixed10m}&\textbf{twolevel10m}&\textbf{fixed100m}&\textbf{twolevel100m}\\
\hline
4 & 1.8080762 & 1.8143984 & 21.56307 & 12.077078 \\
\hline
8 & 1.4035088 & 0.9598364 & 12.79544 & 6.297646 \\
\hline
16 & 1.4310018 & 0.5916096 & 10.666476 & 3.901922\\
\hline
32 & 0.6556778 & 0.404464 & 5.624658 & 2.081128\\
\hline
64 & 0.3043472 & 0.3143486 & 2.946662 & 1.433568\\
\hline
128 & 0.19882468 & 0.30740326 & 1.52265 & 1.392154\\
\hline
\end{tabular}
\vspace{1mm}
\caption{Performance of fixed-size hash tables vs two-level hash tables on AMD Milan\label{tab:tableht1}}
\end{center}
\end{table}

In the two-level implementation, the second level tables were expanded when the number of hash table collisions per slot increased beyond a threshold value~(10 entries) and they were removed when the number of entries fell below this threshold. Different second level tables can be expanded concurrently because of fine-grained read-write locks per first level slot. Both hash table implementations used the memory manager described in the previous section. The two-level implementation used a memory manager per first level slot. This improved page and cache locality in the second level tables. Like in other experiments, we used huge pages and TBB's $malloc$ implementation. The use of read-write locks increased concurrency, by allowing concurrent $finds$ on the same slot using shared read locks and requiring exclusive write locks for $insert$ and $erase$.  

\begin{figure}[!tbph]
\begin{center}	
\includegraphics[width=3.5in,height=2.4in]{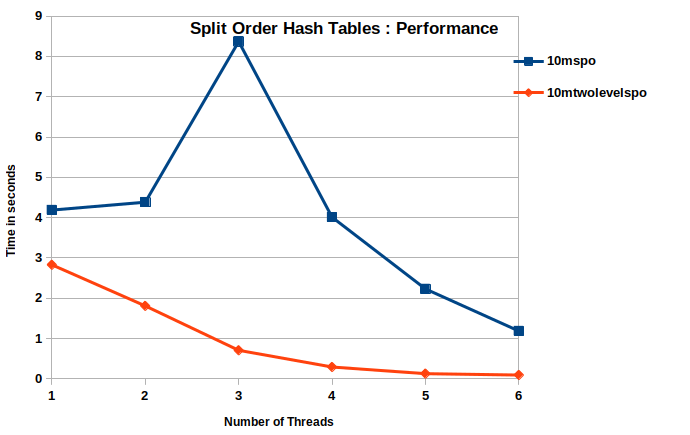}
\caption{Cache behaviour of One-level and Two-level Split-Order~(SPO) Hash Tables on AMD Milan\label{fig:spocache}}
\end{center}
\end{figure}

Split-order hash tables are scalable hash table implementations. It differs from other hash table implementations described in this section because the slots are separate from nodes. Nodes are stored in a singly linked-list in the sorted order of their reverse keys~(LSB to MSB). We fixed an initial number of slots~(seed) and doubled the number of slots whenever occupancy was higher than a threshold value. The maximum number of collisions per slot and seed value are variables for the implementation. For $n$ slots and $m$ maximum collisions per slot, the occupancy of the hash table was computed as $n*m$. When the occupancy $> n*m$, $n$ was increased to $2*n$. The new slots were filled during subsequent hash table operations. When a hash table operation found an empty slot, it was filled by recursively traversing same slots in prior allocations, until a non-empty slot was found. Same slots in previous allocations were located using bit masks. When a non-empty slot was found, the linked-list was split by inserting a dummy node in the list where keys differ in their MSB position as described in~\cite{splitorder}. For example, for $seed=512$ and $n=512$, when the number of slots was increased to $1024$, the $9^{th}$ bit~($log(512)$) was used to identify the split location, where bits are numbered from $0-63$. Splitting performed the required rehashing without data migration. The seed slots were initialized with dummy nodes. Although the implementation discussed in~\cite{splitorder} is lock-free, our implementation used read-write locks on the full hash table as well as per slot. In our implementation, a large slot vector was allocated during initialization. Although exclusive write locks are required, a resizing operation doubles the number of slots and exits, which is a low-cost operation.
The default split-order hash table did not perform well due to cache misses and page faults during rehashing. An empty slot had to recursively access prior slots which caused overheads from cache misses for slots that are far apart. A two-level hierarchical split-order hash table with small number of seed slots  and resizing operations performed per table showed better spatial locality and strong scaling at large thread counts for these workloads. A comparison of overheads caused by cache misses are presented in the graph in figure~\ref{fig:spocache} and tabulated in table~\ref{tab:spocache}. The maximum number of collisions for both split-order implementations was $16$ entries. The seed for split-order hash table was $8192$. For hierarchical split-order hash tables, we used $256$ split-order tables at the first level with seed size $64$ at the second level. Like the other two implementations, the split-order tables were also partitioned across NUMA nodes. We used a single memory manager for the default split-order implementation and a memory manager per first level table for the two-level version. 

\begin{table}[!tbph]
\scriptsize	
\begin{center}	
\begin{tabular}{|c|c|c|}
\hline
\textbf{\#threads} & \textbf{10mspo} & \textbf{10mtwolevelspo} \\
\hline
4 & 4.1893104 & 1.8829426 \\
\hline
8 & 4.384854 & 0.9649104\\
\hline
16 & 8.3696894 & 0.4804762\\
\hline
32 & 4.0107974 & 0.242256\\
\hline
64 & 2.2309622 & 0.1543608\\
\hline
128 & 1.18745908 & 0.11367386\\
\hline
\end{tabular}
\vspace{1mm}
\caption{Cache performance of one-level and two-level split-order~(SPO) hash tables on AMD Milan\label{tab:spocache}}
\end{center}
\end{table}

\begin{table}[!tbph]
\scriptsize
\begin{center}
\begin{tabular}{|c|c|c|c|}
\hline
\textbf{\#threads}&\textbf{100mtbb}&\textbf{100mSPO}&\textbf{100mBinLists}\\
\hline
4 & 7.87826 & 13.57318 & 12.09342 \\
\hline
8 & 4.877724 & 7.092238 & 6.04725 \\
\hline
16 & 4.44002 & 4.032536 & 5.567374 \\
\hline
32 & 2.234972 & 1.890784 & 2.556356 \\
\hline
64 & 1.360036 & 1.124712 & 1.265442 \\
\hline
128 & 0.8601906 & 0.7902118 & 0.6457664 \\
\hline
\end{tabular}
\vspace{1mm}
\caption{Performance of three hash tables~(TBB,Split-Order and BinaryTreeLists) 100m ops on AMD Milan\label{tab:threeht100}}
\end{center}
\end{table}

\begin{table}[!tbph]
\scriptsize
\begin{center}
\begin{tabular}{|c|c|c|c|}
\hline
\textbf{\#threads}&\textbf{1btbb}&\textbf{1bSPO}&\textbf{1bBinLists}\\
\hline
4 & 94.07204 & 165.8882 & 213.8314 \\
\hline
8 & 55.35936 & 84.47286 & 109.2326 \\
\hline
16 & 48.3085 & 44.83896 & 65.62332 \\
\hline
32 & 24.04664 & 22.69882 & 31.12086 \\
\hline
64 & 11.55592 & 11.0454 & 15.21968 \\
\hline
128 & 6.001542 & 5.177758 & 7.701186 \\
\hline
\end{tabular}
\vspace{1mm}
\caption{Performance of three hash tables~(TBB,Split-Order and BinaryTreeLists) 1b ops on AMD Milan\label{tab:threeht}}
\end{center}
\end{table}

\begin{figure}[!tbph]
\begin{center}	
\includegraphics[width=3.5in,height=2.5in]{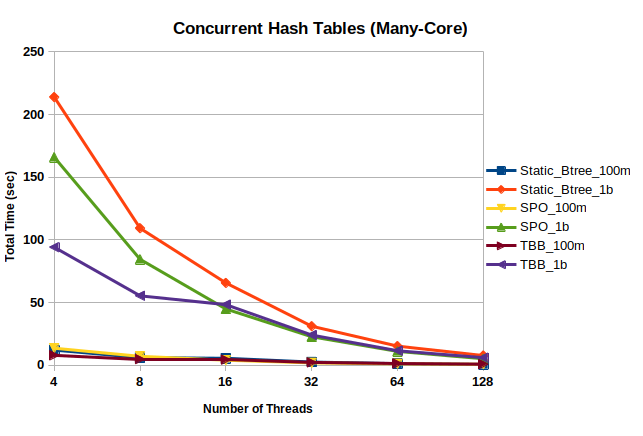}
\caption{Comparison of Three Concurrent Hash Tables~(TBB,Split-Order and BinaryTree lists) on AMD Milan\label{fig:hts}}
\end{center}
\end{figure}

\begin{table}[!tbph]
\scriptsize
\begin{center}
\begin{tabular}{|c|c|c|}
\hline
\textbf{\#threads}&\textbf{100mSPO}&\textbf{1bSPO}\\
\hline
4 & 15.44906 & 154.642 \\
\hline
8 &8.654564 & 85.41856 \\
\hline
16 & 5.51451 & 55.08904 \\
\hline
32 & 3.613338 & 35.90394 \\
\hline
48 & 2.730522 & 27.40854 \\
\hline
\end{tabular}
\vspace{1mm}
\caption{Two-level SPO performance for 100m and 1b operations on Fujitsu A64FX\label{tab:htfu}}
\end{center}
\end{table}

\begin{figure}[!tbph]
\begin{center}
\includegraphics[width=3in,height=2in]{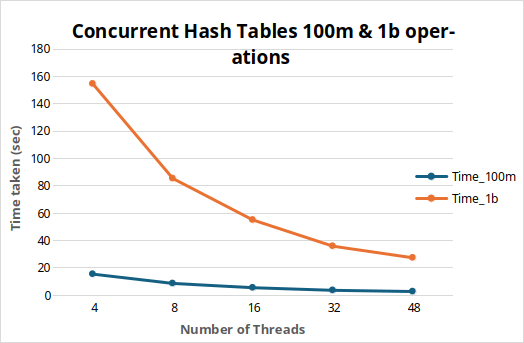}
\caption{Two-level SPO hash table performance for 100m and 1b ops on Fujitsu A64FX\label{fig:htsfu}}
\end{center}
\end{figure}


From our experiments, hierarchical hash tables have higher page and cache hits due to spatial and temporal localities. For the remaining experiments we used two-level tables with binarytreelists and two-level split-order tables. We compared our implementations with the TBB concurrent hash tables~\cite{TBB}. The TBB implementation is similar to a two-level split-order table with expansion and shrinking. Unlike the split-order algorithm, rehashing traverses all entries in a slot, removes and adds them to new slots.  

A comparison between the hash implementations for $100m$ and $1b$ workloads are presented in the graphs in figures~\ref{fig:hts},\ref{fig:htsfu} and tabulated in tables~\ref{tab:threeht100},\ref{tab:threeht} and \ref{tab:htfu}. All observations reported in this section are averaged over $5$ repetitions of each experiment. Both the two-level split-order tables perform better than the two-level binarytreelist for large workloads. The two-level SPO hash-table implementation also showed strong scaling performance on Fujitsu A64FX. The experiments on A64FX are tabulated in the table~\ref{tab:htfu} and plotted in the graph in figure~\ref{fig:htsfu}. For the implementation on A64FX, we used $16$ hash tables per NUMA node at the first level and a seed table size of $4096$. Our implementation is comparable to the TBB implementation in performance, which is a two-level split-order concurrent hash table. The differences in performance between TBB and our implementations at low thread counts are due to differences in memory allocation and the use of micro-locks in TBB. TBB allocates large segments of memory before running hash table queries which leads to improved performance at low thread counts.
All three hash table algorithms are scalable with increasing thread counts. Depending on the available memory and the workload size, a programmer may choose any one of them. Our two-level split-order table is not lockfree. It may be useful to consider a lockfree implementation for comparison.  

\section{Related Work\label{rel}}

The concurrent random skiplist~\cite{concurrentrandomskiplist} from Herlihy et.al is a scalable design with a lock-free \emph{Find} implementation. It requires acquiring locks at all levels for a node and its predecessor for \emph{Addition} and \emph{Deletion} operations. Our 1-2-3-4 tree implementation performs more computation per operation for maintaining the balance criterion. The concurrent skiplist implementation provided by Java is based on lock-free linked-lists~\cite{harris}. The contention in our design can be reduced using a lock-free implementation and re-balancing operations similar to those described in~\cite{lockfreebtree}. There are several scalable lock-free designs of concurrent binary search trees~(BST)~\cite{lockfreeBSTnag},~\cite{lockfreeBSTellen},~\cite{ellen},~\cite{lockfreeAVL}. Some of these BST implementations have included lock-free re-balancing operations, but with relaxations in the BST definition or balance criteria. A survey of concurrent BST performance can be found in~\cite{lockfreebstperf}. Skiplists are more convenient than binary search trees for range searches because of the terminal linked-list. All keys that lie within a range can be found by locating the minimum key and following pointers whereas a binary tree will have to perform depth-first traversal on sub-trees.

LCRQ~\cite{lcrq} is one of the fastest concurrent unbounded queues available. Algorithms for concurrent bounded length lock-free queues have shown scalability on GPUs~\cite{broker}. Bounded length queue implementations work well for workloads which have alternate \emph{push} and \emph{pop}. They also have better page and cache hits from limiting the buffer size. We did not follow that direction because we did not want to make assumptions regarding our workloads. 

Concurrent cuckoo~\cite{cuckoo} and hopscotch~\cite{hopscotch} hash table implementations have shown better performance than TBB's implementation for fixed size tables and low occupancy. Their performance gains come from higher concurrency due to fine-grained locks per slot and better spatial and temporal localities that leads to higher cache hits.  
Other widely used concurrent hash tables are those that perform open-addressing. These implementations have higher concurrency and are lock-free. But we did not consider open-addressing hash tables in this paper because they have potential problems from clustering which affect the computational complexity of table operations. The expected cost of search operations can be as high as $\frac{\pi N}{2}^{\frac{1}{2}}$ for nearly full tables with linear probing~\cite{sedgewickanalysis}. This is expensive for the workload sizes discussed in this paper. A detailed discussion of different types of hash tables and hash functions can be found in ~\cite{gonnet}. Deletion in concurrent open-addressing designs are performed lazily by marking deleted entries. \cite{openaddr1} discusses open-addressing implementations that have scaled well for fixed size tables. However, they have not discussed performance for workloads which require frequent resizing, especially with an implementation that supports lazy deletion.     

Data structure implementations optimized for NUMA architectures have been discussed in ~\cite{numaperf}. They have used redundancy and combiners to reduce memory accesses from remote nodes. But the authors of article~\cite{numaperf} have not discussed the correctness of their implementation, especially since the different NUMA nodes will have different views of the same set of global operations. In our experiments, we used lock-free queues for inter-node communication and found that it does not affect the scalability of the program in spite of memory accesses from remote NUMA nodes for \emph{push} operations. Although we used two data structures, i.e a lock-free queue and skiplist or hash table, we allocated separate memory blocks for them which provided isolation and spatial and temporal localities in our programs. In our experiments, we filled the queues first before performing operations on the data structures which is also a reason for good locality. Programs which fill queues while performing operations on data structures may also be interesting experiments for this design. 
\section{Conclusions and Future Work\label{conc}}

This paper discusses the design of a concurrent deterministic 1-2-3-4 skiplist. We found the implementation to be scalable on many-core nodes. To the best of our knowledge, this is the first concurrent implementation of a deterministic skiplist. This implementation has guaranteed $O(logn)$ cost for all operations, unlike random skiplists which have variable number of links. It is easier to analyse the performance of skiplists and compare them  with BSTs or red-black trees using a deterministic implementation. We have also provided performance results for two other widely used data structures on many-core nodes. We found non-NUMA memory accesses and page faults from remote NUMA nodes to be the most expensive overheads on AMD Milan~\cite{AMD} as well as Fujitsu A64FX~\cite{fujitsu}. We used aggressive compiler optimizations for both processors. Compiler optimizations such as simd vectorization, prefetching and loop unrolling improved performance on both architectures. We discussed methods for memory management in many-core nodes in this paper. In future, we plan to port these implementations to GPUs which have higher concurrency and different types of memory latencies compared to CPUs. These implementations can be easily made distributed by adding another level of partitioning and interprocess communication via MPI or RPCs. Since the implementations are correct and linearizable at process level, the overall correctness of the distributed implementation is guaranteed. 

\bibliographystyle{IEEEtran}
\bibliography{ref}

\end{document}